\documentclass{ws-ijqi}

\usepackage[super,sort&compress]{natbib}

\newcommand{\etal}{{\it{et al.~}}}
\newcommand{\Tr}{\mathrm{ Tr }}
\newcommand{\ie}{{\it{i.e.~}}}

\newcommand{\bra}[1]{\langle #1 |}
\newcommand{\ket}[1]{| #1 \rangle}

\def\one{{\mathchoice {\rm 1\mskip-4mu l} {\rm 1\mskip-4mu l} {\rm
1\mskip-4.5mu l} {\rm 1\mskip-5mu l}}}

\begin{document}

\markboth{S. Boixo, L. Aolita, D. Cavalcanti, K. Modi, and A. Winter}
{Quantum locking of classical correlations and  quantum discord of classical-quantum states }

\catchline{}{}{}{}{}

\title{Quantum locking of classical correlations and  quantum discord of classical-quantum states}

\author{S. Boixo}
\address{Department of Chemistry and Chemical Biology, Harvard University,\\ 
Cambridge MA, USA\\ 
boixocastrillo@fas.harvard.edu}

\author{L. Aolita}
\address{ICFO-Institut de Ci\`{e}ncies Fot\`{o}niques, Mediterranean
Technology Park,\\ 08860 Castelldefels (Barcelona), Spain} 

\author{D. Cavalcanti}
\address{Centre for Quantum Technologies, National University of Singapore, Singapore}

\author{K. Modi}
\address{Centre for Quantum Technologies, National University of Singapore, Singapore}


\author{A. Winter}
\address{Department of Mathematics, University of Bristol, Bristol, UK\\
Centre for Quantum Technologies, National University of Singapore, Singapore}

\maketitle

\begin{history}
\end{history}

\begin{abstract}
  A locking protocol between two parties is as follows: Alice gives an
  encrypted classical message to Bob which she does not want Bob to be
  able to read until she gives him the key. If Alice is using
  classical resources, and she wants to approach unconditional
  security, then the key and the message must have comparable sizes. But if
  Alice prepares a quantum state, the size of the key can be comparatively
  negligible. This effect is called quantum locking. Entanglement does not
  play a role in this quantum advantage. We show that, in this
  scenario, the quantum discord quantifies the advantage of the
  quantum protocol over the corresponding classical one for any
  classical-quantum state.
\end{abstract}


\section{Introduction}	
The separation between classical and quantum correlations in a quantum system has puzzled physicists since the early days of quantum information science~\cite{werner_quantum_1989,zurek2000,ollivier,henderson,groisman,li_total_2007,kaszlikowski_quantum_2008,bennett_postulates_2011,luo_using_2008,wu_correlations_2009}. The fact that \emph{unentangled} or \emph{separable} states can be created by local operations and classical communication~\cite{werner_quantum_1989} lead to the belief that all quantum correlations which are non-classical can be ascribed to entanglement. A bipartite quantum state is separable or unentangled if it can written in the form
\begin{align}\label{eq:separable}
 \rho_{AB}  = \sum_j p_j \rho_A^{(j)} \otimes \rho_B^{(j)}\;.
\end{align}
The subscripts $A$ and $B$, or Alice and Bob, denote the two parties. In fact, all correlations of separable quantum states are local, {\it i.e.} can be explained by classical local-hidden-variable models \cite{werner_quantum_1989}. Nevertheless, Ollivier and Zurek gave a new measure of non-classical correlations
according to which some separable quantum states can contain quantum correlations~\cite{zurek2000,ollivier}. For separable states, this comes simply from the possibility of choosing among different non-orthogonal local bases - a typical quantum feature - in the preparation and measurement. The measure proposed by Ollivier and
Zurek to quantify these more general quantum correlations is called \emph{quantum discord}~\cite{zurek2000,ollivier}.



Quantum discord has received an astonishingly amount of interest recently~\cite{wu_correlations_2009,dattashaji,datta_signatures_2009,ferraro,werlangPRA,Maziero09,piani,CesarEtal07,shabani_vanishing_2009,sarandy_classical_2009,maziero_quantum_2010,luo_quantum_2008,modietal,paris10,Adesso10,mazzola,fanchini,dvb10,datta10}, triggered mostly by the possibility of being the reason for the speedup of a quantum computation model~\cite{dattashaji} (called deterministic quantum computation with one qubit, or DQC1 for short~\cite{knill_power_1998}). In the last years, quantum discord has been studied in several contexts like quantum computation~\cite{dattashaji,datta_signatures_2009}, decoherence processes~\cite{ferraro,werlangPRA,Maziero09}, local broadcasting~\cite{piani}, open-quantum-system formalism~\cite{CesarEtal07,shabani_vanishing_2009}, and quantum phase transitions~\cite{sarandy_classical_2009,maziero_quantum_2010}, just to cite some. Recently, more formal characterizations of quantum discord have been put forward~\cite{ferraro,luo_quantum_2008,modietal}, and it has also been probed in the laboratory~\cite{Lanyon,jsxu}. Finally, quantum discord can be related to the the difference between the efficiency of quantum and classical Maxwell's 
demons~\cite{zurek_quantum_2003,brodutch_quantum_2010} and it has been given information theoretical operational interpretations as the entanglement consumption in an extended quantum state merging protocol~\cite{cavalcanti_operational_2011} and as the markup in the cost of quantum communication in the process of quantum state merging~\cite{madhok_interpreting_2011}.

\section{Quantum discord and accessible information}
We review now the definition of quantum discord. A good place to start is by reminding ourselves of the information theoretical classical measure of correlations for classical random variables. We associate temporarily to each party Alice and Bob a classical random variable, denoted $A$ and $B$ respectively. Then the amount of correlations in the classical probability distribution $p(A,B)$  can be measured by the mutual information
\begin{align}
  I(A ; B) = D_{\rm KL} (p(A,B) \parallel p(A) p(B) )\;,
\end{align}
where $p(A)$ and $p(B)$ are the marginal distributions and $D_{\rm KL}$ is the Kullback-Leibler divergence which measures the distance or relative entropy. That is, the mutual information is the distance to the corresponding uncorrelated probability distribution. An equivalent expression for the mutual information is found in terms of the (classical) conditional entropy, which itself has the expression
\begin{align}\label{eq:classical_conditional_entropy}
 H(A|B) = \sum_b p(b) H(A|B=b) = H(A,B) - H(B)\;,
\end{align}
where $H = - \sum_b p(b) \log p(b)$ is the Shannon entropy. The (classical) conditional entropy is the expected value of the entropies of the conditional distribution. The mutual information gets the expression
\begin{align}
 I(A ; B) = H(A) - H(A|B) = H(A) + H(B) - H(A,B)\;, 
\end{align}
The mutual information is then the reduction in the uncertainty of Alice random variable due to the knowledge of Bobs outcome~\cite{cover_elements_2006}.

Quantum states $\rho_{AB}$ give rise to similar expressions when substituting each Shannon entropy $H(A)$ by the von Neumann entropy $S(\rho_A) = - \Tr \rho_A \log \rho_A$. In what follows we will  use the notation $S(A) = S(\rho_A)$ and so for. The quantum mutual information is
\begin{align}\label{eq:mutual_information}
 I(A ; B) = S(A) - S(A|B) = S(A) + S(B) - S(A,B)\;,
\end{align} 
which is also a measure of the total amount of quantum correlations. Indeed, the quantum mutual information is also the relative entropy between $\rho_{AB}$ and $\rho_A \otimes \rho_B$. Operationally, it can be seen that it corresponds to the minimal rate 
of randomness that is required to completely erase all the 
correlations in $\rho_{AB}$~\cite{groisman}.


Consider now the following scenario, quite standard in a quantum communication context: Alice produces classical (orthogonal) states $\ket{a}\bra{a}$ according to a probability distribution $\{p_a\}$ and communicates with Bob through a quantum channel. Bob wants to know  the value or ``letter'' $a$ that Alice holds. If the transmission channel is perfect, Bob can reliably determine which $\ket{a}\bra{a}$ was sent. However, if the channel used to transmit the states introduces errors, Bob will usually get mixed states $\sigma_B^{(a)}$.  In this scenario the final quantum state shared between Alice and Bob in the enlarged Hilbert space representation is a classical-quantum state, given by
\begin{equation}\label{cq state}
\rho_{cq}=\sum_a p_a \ket{a}\bra{a}_A\otimes\sigma_B^{(a)}.
\end{equation}
Notice that, because Alice holds a classical system, there is no entanglement involved in this quantum effect. All classical-quantum states are of the form given by Eq.~\eqref{eq:separable}.

Assume that Bob now measures his state to obtain information about Alice's letter.
The maximum amount of \emph{classical correlations} that Bob can obtain about Alice's letter in the previous scenario is quantified by the accessible information $I_{\rm acc}$~\cite{nielsen}. This is defined as the maximum  mutual information $I(A;B)$ that he can extract from $\rho_{cq}$ by making a measurement $M_B$, \ie
\begin{align}
I_{\rm acc}(\rho_{AB})&=\max_{M_B} I(A;B)\nonumber\\
&=S(\rho_A)-\min_{M_b} \sum_b p_b S(\rho_{A|b}),\label{eq:accessible_information}
\end{align}
where the state $\rho_{A|b}$ is the state of Alice given that Bob
performs a measurement $M_B=\{M_b\}$ in his subsystem and receives
outcome $b$, \ie $ \rho_{A|b}=\Tr_B[M_b \rho] /p_b$.  The probability
of outcome $b$ is given by $p_b=\Tr [\rho M_b]$. Notice that the
expected values of the entropies is the conditional information, in
accordance with Eq.~\eqref{eq:classical_conditional_entropy}.

At this point, a natural way of defining purely \emph{quantum correlations} in a quantum state is to subtract the amount of classical correlations accessible to Bob from the total amount of correlations that the state contains. This is precisely how quantum discord is defined: the quantum discord of a bipartite quantum state $\rho_{AB}$ is 
\begin{equation}
D(\overleftarrow{AB})= I(A;B)-I_{\rm acc}(\rho_{AB}).
\label{disco}
\end{equation}

The corresponding general version of the accessible information (with the same definition, but Alice's state is not assumed classical) has been proposed as a measure of classical correlations~\cite{henderson}. Quantum discord can also be defined for all quantum states with the same expression~\eqref{eq:accessible_information}~\cite{zurek2000,ollivier} (not just classical-quantum states).  Interestingly, part of the motivation of the original definition of quantum discord is that the quantum conditional entropy used in Eq.~\eqref{eq:mutual_information} can take negative values, whereas the average conditional entropy of Eq.~\eqref{eq:accessible_information} is always positive. Later the negative values of the quantum conditional information was given an operational interpretation in the quantum state merging protocol, and this plays a key role in the later operational interpretations of quantum discord~\cite{cavalcanti_operational_2011,madhok_interpreting_2011}.


\section{Quantum locking of classical correlations}
Before stating our main result let us describe the idea of locking classical correlations in quantum systems. We first present the simplest example of this protocol, and calculate the quantum discord in this case. This will help gain intuition and understand better the more general case that we will present later. We should remark that the connection between the quantum discord and quantum locking of classical correlations was suggested in Refs.~\cite{datta_signatures_2009, wu_correlations_2009}. Here we will prove that this connection is indeed valid for classical-quantum states.

The protocol is now this: Alice wants to give a message $a$  to Bob, but wants to keep it secret for the moment. She encrypts the message, and does not want Bob to able to read it until she gives him the key $k$. Think of the encrypted mission plans of a war movie. We are interested in unconditional security which, barring a security proof of a more practical encryption protocol, can only mean information theoretical security. 

Our aim is to minimize the size of the key. Continuing with the war time analogy, it is very expensive to secure a channel of communication to deliver the key after Bob departs, and every bit counts. How big must the key be if we are limited to using classical resources? Let $m$ denote the number of bits of Alice's message.  Bob must be able to read the message after receiving the key, and therefore the mutual information between the message and Bob's quantum state plus key must be $m$ 
\begin{align}
 I( A ; B,K) = m \;,
\end{align}
where $K$ is the random variable corresponding to the keys. Using the chain rule for conditional entropies
\begin{align}
 m = I(A;B,K) = I(A;B) + I (A;K|B) \;.
\end{align}
The conditional mutual information is defined as the mutual information but using conditional entropies, and the chain rule follows directly from this definition. If Alice wants to approach unconditional security then $I(A;B)$ must be negligible. In addition, because at this point we are assuming that Alice and Bob are using classical resources~\footnote{In fact, it is enough that the key is classical. To see this, we can first use the chain rule to write
\begin{align*}
  I(A;K|B) = I(AB;K) - I(B;K)\;.
\end{align*}
The term $I(B;K)$ is positive by subadditivity. The first term, in the general (quantum) case is
\begin{align*}
  I(AB;K) = S(K) - S(K|B) \le |K|\;.
\end{align*}
The right hand size inequality follows because the conditional entropy is non-negative, given that there is no entanglement.}, the conditional mutual information $I(A;K|B)$ is bounded by the size of the key $|K|$. 
Therefore, the key must be almost as big as the message $m \sim |K|$. For example, in one-time pad a message of $m$ bits is added to a random key of the same length, which hides the message in the (classical) correlations between the key and the message.

The fact that for classical locking of correlations with almost unconditional security we must have $m \sim |K|$ is consistent with the following intuition: one should expect that by the transmission of $l$ bits Alice and Bob cannot increase the correlations by more than $l$ bits. In fact, this is basically the content of the principle of Information Causality, which can be used to explain the Tsirelson bound of the CHSH inequality~\cite{pawlowski_information_2009}. This principle basically follows from the chain rule of conditional entropies, as in the derivation above.

The chain rule is indeed obeyed by several measures of correlations, such as the quantum and classical mutual information. Surprisingly, DiVincenzo \etal showed that the accessible information $I_{\rm acc}$ violates this rule~\cite{divincenzo_locking_2003}: some states have an arbitrarily large amount of classical correlations unlocked after the exchange of some small amount (even one bit) of communication. This effect became known as \emph{quantum locking of classical correlations}. In other words, the transmission of $l$ bits results in a much larger increase of the accessible correlations.

We can define the amount of extra correlations as
\begin{equation}\label{delta}
\Delta=I_{\rm acc}(A,K;B,K)-(I_{\rm acc}(A,K;B)+|K|).
\end{equation}
This quantity can be interpreted as follows. The first term, $I_{\rm acc}(A,K;B,K)$, is the maximum amount of correlations Bob can get if he waits for the communication of the key from Alice to make the measurement in his system. The second term, $I_{\rm acc}(A,K;B)+|K|$, refers to the maximum amount of correlations that Bob can get if he measures before receiving the key. Thus, $\Delta$ can be thought as the amount of extra  correlations that Bob can get if he waits for Alice's communication due to the quantum unlocking of classical correlations.

In the simplest case, the state considered by DiVincenzo \etal is:
\begin{equation}\label{locking state}
\rho_{AB}=\frac{1}{2^{m+1}}\sum_{a=0}^{2^m-1}\sum_{k=0}^1 \ket{a,k}\bra{a,k}_A\otimes(U_k \ket{a}\bra{a}U_k^\dagger)_B,
\end{equation}
being $U_0=\one$ and $U_1$ a unitary such that $U_1\ket a$ is a mutually unbiased basis with respect to $\ket a$. The accessible information without the key $I_{\rm acc}(A,K;B)$ is $m/2$ bits~\cite{divincenzo_locking_2003}, while the accessible information with the key $I_{\rm acc}(A,K;B,K)$ is $m+1 $. The excess of accessible information $\Delta$ unlocked with the single bit $k$ is $m/2$. The quantum mutual information $I(A,K;B)$ is $m$. The difference between the quantum mutual information and the accessible information without the key, that is, the discord $D(\overleftarrow{AB})$ of the initial state~\cite{wu_correlations_2009}, is also $m/2 = \Delta$. Therefore, the discord of this state, without entanglement, describes the advantage for quantum locking, the difference between the optimal classical increment of shared information, and the one achievable in this protocol
.
  There exist quantum locking protocols that have an arbitrary small amount of prior accessible information, and the rest is unlocked by a small key~\cite{divincenzo_locking_2003,hayden_randomizing_2004,fawzi_low-distortion_2010}.


\section{Locking and discord of general classical-quantum states} 
We have just shown that the quantum discord quantifies the amount of extra correlations that Alice and Bob gets if Bob waits until he gets some communication to perform a measurement, if they share the state~\eqref{locking state}. As we proceed to show now, the same interpretation is valid for more general classical-quantum states \eqref{cq state}. The only difference will be that, in general, Alice and Bob will have to share many copies to attain the quantities $I_q$ and $I_{\rm acc}$. In other words, we will work in the so-called asymptotic regime.

Let us now show that for any locking protocol that, by design, works with a single copy~\cite{divincenzo_locking_2003,hayden_randomizing_2004,fawzi_low-distortion_2010}, the amount of quantum locked correlations is always equal to its quantum discord.
Note that, since after receiving the key the message $a$ is completely revealed to Bob we must have
\begin{equation}
I_{\rm acc}(A,K;B,K)=I_q(A,K;B,K)=m+|K|.
\end{equation} 
On the other hand, 
\begin{equation}
I_q(A,K;B,K)\leq I_q(A,K;B)+|K|\leq m+|K|.
\end{equation} 
As explained above, the first inequality comes from the fact that the quantum mutual information can not increase by more than the $|K|$ bits of information being transmitted, while the second is simply a bound on the correlations of the encrypted message $|B|=m$. We thus conclude that
\begin{equation}
I_{\rm acc}(A,K;B,K)=I_q(A,K;B,K)= I_q(A,K;B)+|K|.
\end{equation}
Finally, plugging it back in \eqref{delta}, we get
\begin{equation}\label{delta2}
\Delta=I_q(A,K;B)-I_{\rm acc}(A,K;B)= D(\overleftarrow{AB})
\end{equation}
which is nothing but the quantum discord.

For the general classical-quantum state we point out that asymptotically (in the limit of many copies), the discord of a CQ state is always equal to the advantage of quantum locking. This can be seen from the classical data compression with quantum side information result of Ref.~\refcite{devetak_classical_2003}, which builds upon the \emph{product} classical information capacity of a noisy quantum channel (HSW)~\cite{holevo_bounds_1973,schumacher_sending_1997}. We only give the intuition here and refer to Ref.~\refcite{devetak_classical_2003} and references therein for the details.

The intuition for the HSW result is the same as for
Shannon's noisy channel coding theorem. Alice builds a quantum code by randomly choosing a subset of size $2^{nR}$ of strings of size $n$ generated according to some distribution $A$. For each input letter $a$, Bob gets output $\sigma_B^{(a)}=\mathcal E(\rho_a)$, where $\rho_a$ is part of the quantum code construction. The rate of the code is $R$. Denote by $a^n$ a possible input of size $n$, and by $\sigma_B^{({a^n})}$ the corresponding output. The dimension of the subspace of typical sequences for the output of the channel on Bob's size goes like $2^{nS(B)}$. In average, the dimension of the subspace corresponding to the conditional typical output sequences (taking into account the channel noise) goes like $2^{n S(B|A)}$. That is, in average, the number of outputs for each 
input is $2^{nS(B|A)}$. The size of the subset of inputs in the quantum code, given by the rate $R$, must be such that we can identify the code word corresponding to the output, so we must assign a subspace of dimension $2^{n S(B|A)}$, out of the total $2^{nS(B)}$, for each code word. That is, the rate must obey
\begin{align}
  2^{nR} &\approx 2^{n S(B)}/2^{n S(B|A)} = 2^{n(S(B)-S(B|A))} \notag \\ &= 2^{n I(A;B)}\;.
\end{align}
This is the content of the HSW theorem: the product classical information capacity of a noisy quantum channel is given by the mutual information $I(A;B)$.

With this background, we can approach the content of Ref.~\refcite{devetak_classical_2003}. The
question is, for a CQ state, how many quantum codes are necessary to
cover the typical sequences of $n$ copies? There are $2^{n S(A)}$
typical sequences in Alice's side, and a quantum code, by the HSW theorem, has
size $2^{n I(A;B)}$, so the number of necessary quantum codes is $2^{n(S(A)-I(A;B))}=2^{n S(A|B)}$. Then Alice sends only information identifying the quantum code, $S(A|B)$ bits per copy of the CQ state,
and Bob measures with that code, obtaining Alice's complete message. 

When Bob is forced to measure each copy of the output state before Alice sends the key, the amount of information unknown to Bob is quantified by Alice's entropy conditional on Bob's measurement outcome, that is, $\min_{M_B} \sum_b p_b S(\rho_{A|b})$. Because Bob measurement turns the quantum state into a classical random variable, the optimal choice of measurement for the accessible information also defines a corresponding classical protocol. The number of bits that Alice must send to Bob in this case is $\min_{M_B} \sum_b p_b S(\rho_{A|b})$. We have seen that, using quantum codes, Alice must send only $S(A|B)$ bits. The difference is the quantum advantage for quantum locking, quantified by the discord $D(\overleftarrow{AB})$ for CQ states
\begin{align}
  D(\overleftarrow{AB}) = I(A;B)-I_{\rm acc}(A;B) = \min_{M_B} \sum_b p_b S(\rho_{A|b}) - S(A|B)\;.
\end{align}

\section{Conclusions}
For a classical-quantum state, the difference between the accessible correlations and the quantum correlations gives a measure of purely quantum correlations, called quantum discord. This is despite the fact that the quantum state is separable and, therefore, admits a local hidden variable description. These purely quantum correlations are the source of the quantum advantage of a quantum locking protocol. In the simplest case, Alice encrypts her message in the choice of basis (as in the BB84 cryptographic protocol~\cite{bennett_quantum_1984}). If Alice then announces the choice of basis to Bob, he can perform a measurement which unambiguously identifies the encrypted message. For a general classical-quantum state, the state $\rho_{AB}^{\otimes n}$ also has the property of being decomposable into subensembles with mutually orthogonal elements in the asymptotic limit. In both cases, quantum discord quantifies the quantum advantage. This also points towards a relation of discord to some quantum cryptography protocols. 

\section*{Acknowledgments}
Part of this work was done while SB was at the Institute for Quantum Information at the California Institute of Technology. This work was supported by the National Research Foundation, the Ministry of Education of Singapore, the Spanish ``Juan de la Cierva" Programme, the Royal Society, U.K. EPSRC, the European Commission, ERC and the Philip Leverhulme Trust. 

\bibliographystyle{ws-ijqi}
\bibliography{omd}

\begin{thebibliography}{10}

\bibitem{werner_quantum_1989}
R.~F. Werner, {\em Phys. Rev. A} {\bf 40} (October 1989) p. 4277.

\bibitem{zurek2000}
W.~H. Zurek, {\em Annalen der Physik (Leipzig)} {\bf 9}  (2000) p. 853.

\bibitem{ollivier}
H.~Ollivier and W.~Zurek, {\em Phys. Rev. Lett.} {\bf 88} (Jan 2001) p. 017901.

\bibitem{henderson}
L.~Henderson and V.~Vedral, {\em J. Phys. A: Math. Gen.} {\bf 34}  (2001) 6899.

\bibitem{groisman}
B.~Groisman, S.~Popescu and A.~Winter, {\em Phys. Rev. A} {\bf 72} (Sep 2005)
  p. 032317.

\bibitem{li_total_2007}
N.~Li and S.~Luo, {\em Phys. Rev. A} {\bf 76}  (2007) p. 032327.

\bibitem{kaszlikowski_quantum_2008}
D.~Kaszlikowski, A.~{Sen(De)}, U.~Sen, V.~Vedral and A.~Winter, {\em Phys. Rev.
  Lett.} {\bf 101}  (2008) p. 070502.

\bibitem{bennett_postulates_2011}
C.~H. Bennett, A.~Grudka, M.~Horodecki, P.~Horodecki and R.~Horodecki, {\em
  Phys. Rev. A} {\bf 83} (January 2011) p. 012312.

\bibitem{luo_using_2008}
S.~Luo, {\em Phys. Rev. A} {\bf 77} (February 2008) p. 022301.

\bibitem{wu_correlations_2009}
S.~Wu, U.~V. Poulsen and K.~M{\o}lmer, {\em Phys. Rev. A} {\bf 80}  (2009) p.
  032319.

\bibitem{dattashaji}
A.~Datta, A.~Shaji and C.~Caves, {\em Phys. Rev. Lett.} {\bf 100}  (2008) p.
  050502.

\bibitem{datta_signatures_2009}
A.~Datta and S.~Gharibian, {\em Phys. Rev. A} {\bf 79} (April 2009) p. 042325.

\bibitem{ferraro}
A.~Ferraro, L.~Aolita, D.~Cavalcanti, F.~M. Cucchietti and A.~Acin, {\em Phys.
  Rev. A} {\bf 81}  (2010) p. 052318.

\bibitem{werlangPRA}
T.~Werlang, S.~Souza, F.~F. Fanchini and C.~J. Villas-Boas, {\em Phys. Rev. A}
  {\bf 80}  (2009) p. 024103.

\bibitem{Maziero09}
J.~Maziero, L.~C. Celeri, R.~M. Serra and V.~Vedral, {\em Phys. Rev. A} {\bf
  80}  (2009) p. 044102.

\bibitem{piani}
M.~Piani, P.~Horodecki and R.~Horodecki, {\em Phys. Rev. Lett.} {\bf 100}
  (2008) p. 090502.

\bibitem{CesarEtal07}
C.~A. Rodriguez-Rosario, K.~Modi, A.-M. Kuah, A.~Shaji and E.~C.~G. Sudarshan,
  {\em J. Phys. A: Math. Gen.} {\bf 41}  (2008) p. 205301.

\bibitem{shabani_vanishing_2009}
A.~Shabani and D.~A. Lidar, {\em Phys. Rev. Lett.} {\bf 102} (March 2009) p.
  100402.

\bibitem{sarandy_classical_2009}
M.~S. Sarandy, {\em Phys. Rev. A} {\bf 80}  (2009) p. 022108.

\bibitem{maziero_quantum_2010}
J.~Maziero, H.~C. Guzman, L.~C. C\'{e}leri, M.~S. Sarandy and R.~M. Serra, {\em
  Phys. Rev. A} {\bf 82} (July 2010) p. 012106.

\bibitem{luo_quantum_2008}
S.~Luo, {\em Phys. Rev. A} {\bf 77} (April 2008) p. 042303.

\bibitem{modietal}
K.~Modi, T.~Paterek, W.~Son, V.~Vedral and M.~Williamson, {\em Phys. Rev.
  Lett.} {\bf 104}  (2010) p. 080501.

\bibitem{paris10}
P.~Giorda and M.~G.~A. Paris, {\em Phys. Rev. Lett.} {\bf 105}  (2010) p.
  020503.

\bibitem{Adesso10}
G.~Adesso and A.~Datta, {\em Phys. Rev. Lett.} {\bf 105}  (2010) p. 030501.

\bibitem{mazzola}
L.~Mazzola, J.~Piilo and S.~Maniscalco, {\em Phys. Rev. Lett.} {\bf 104}
  (2010) p. 200401.

\bibitem{fanchini}
F.~F. Fanchini, T.~Werlang, C.~A. Brasil, L.~G.~E. Arruda and A.~O. Caldeira,
  {\em Phys. Rev. A} {\bf 81}  (2010) p. 052107.

\bibitem{dvb10}
B.~Daki\'c, V.~Vedral and v.~Brukner, {\em Phys. Rev. Lett.} {\bf 105} (Nov
  2010) p. 190502.

\bibitem{datta10}
A.~Datta, {\em arXiv:1003.5256}   (2010)

\bibitem{knill_power_1998}
E.~Knill and R.~Laflamme, {\em Phys. Rev. Lett.} {\bf 81} (December 1998) p.
  5672.

\bibitem{Lanyon}
B.~P. Lanyon, M.~Barbieri, M.~P. Almeida and A.~G. White, {\em Phys. Rev.
  Lett.} {\bf 101}  (2008) p. 200501.

\bibitem{jsxu}
J.-S. Xu {\em et~al.}, {\em Nat. Comm.} {\bf 1}  (2010) p.~7.

\bibitem{zurek_quantum_2003}
W.~H. Zurek, {\em Phys. Rev. A} {\bf 67} (January 2003) p. 012320.

\bibitem{brodutch_quantum_2010}
A.~Brodutch and D.~R. Terno, {\em Phys. Rev. A} {\bf 81} (June 2010) p. 062103.

\bibitem{cavalcanti_operational_2011}
D.~Cavalcanti, L.~Aolita, S.~Boixo, K.~Modi, M.~Piani and A.~Winter, {\em Phys.
  Rev. A} {\bf 83} (March 2011) p. 032324.

\bibitem{madhok_interpreting_2011}
V.~Madhok and A.~Datta, {\em Phys. Rev. A} {\bf 83} (March 2011) p. 032323.

\bibitem{cover_elements_2006}
T.~M. Cover and J.~A. Thomas, {\em Elements of information theory} (John Wiley
  and Sons, 2006).

\bibitem{nielsen}
M.~A. Nielsen and I.~L. Chuang, {\em Quantum Computation and Quantum
  Information} (Cambridge University Press, Cambridge, UK, 2000).

\bibitem{pawlowski_information_2009}
M.~Pawlowski, T.~Paterek, D.~Kaszlikowski, V.~Scarani, A.~Winter and
  M.~Zukowski, {\em Nature} {\bf 461} (October 2009) 1101.

\bibitem{divincenzo_locking_2003}
D.~P. DiVincenzo, M.~Horodecki, D.~W. Leung, J.~A. Smolin and B.~M. Terhal,
  {\em Phys. Rev. Lett.} {\bf 92}  (2003) p. 067902.

\bibitem{hayden_randomizing_2004}
P.~Hayden, D.~Leung, P.~W. Shor and A.~Winter, {\em Comm. Math. Phys.} {\bf
  250}  (2004) p. 371.

\bibitem{fawzi_low-distortion_2010}
O.~Fawzi, P.~Hayden and P.~Sen, {\em arXiv:1010.3007}  (October 2010)

\bibitem{devetak_classical_2003}
I.~Devetak and A.~Winter, {\em Phys. Rev. A} {\bf 68}  (2003) p. 42301.

\bibitem{holevo_bounds_1973}
A.~S. Holevo, {\em Probl. Inf. Transm.} {\bf 9}  (1973) p. 177.

\bibitem{schumacher_sending_1997}
B.~Schumacher and M.~D. Westmoreland, {\em Phys. Rev. A} {\bf 56} (July 1997)
  p. 131.

\bibitem{bennett_quantum_1984}
C.~H. Bennett and G.~Brassard, {\em Quantum cryptography: Public key
  distribution and coin tossing}, in {\em Proc. IEEE Int. Conf. Computers,
  systems and Signal Processing\/},  1984, p. 175.

\end{thebibliography}

\end{document}